\documentclass[10pt, draft, showpacs, amsmath,amsfonts, amssymb,
,prd]{revtex4} 
\usepackage{bm}
\usepackage{latexsym}
\usepackage{dcolumn}
\usepackage{amsfonts,amssymb}
\usepackage{graphicx,epsfig}
\usepackage{amsmath}
\begin{document}



\title{Photon propagator for axion electrodynamics}

\author{Yakov Itin}

\affiliation{Institute of Mathematics, Hebrew University of
   Jerusalem \\
   and Jerusalem College of Technology\\
  email: {\tt itin@math.huji.ac.il}}

\begin{abstract}
The axion modified electrodynamics is usually used as a model for description of possible  violation of Lorentz invariance in field theory. The low-energy manifestation of
Lorentz violation can hopefully be observed in experiments with the electromagnetic
waves. It justifies the importance of study  how a small axion addition can modify the  wave propagation.  Although a constant axion  does not contribute to the dispersion relation at all, even a slowly varying axion field destroys  the light  cone structure. 
In this paper, we study the  wave propagation in the axion modified electrodynamics in the framework of premetric approach. 
In an addition to the modified dispersion relation, we derive the axion generalization of the photon propagator in Feynman and Landau gauge.  Our consideration is free from the usual restriction of a usual restriction to the constant gradient axion field. It is remarkable   that the axion modified propagator is hermitian. Consequently the dissipation effects  absent even in the phenomenological model considered here.   
\end{abstract}
\date{\today}
  \maketitle
{\bf 1. Axion electrodynamics in premetric formalism.}
  Although  Lorentz invariance is a basic assumption of the standard classical
and quantum field theory,  this invariance
is probably violated in  quantum gravity and string theory. 
One believes that the low-energy manifestation of
Lorentz violation can be observed in experiments with the ordinary electromagnetic
waves. 
  Axion electrodynamics \cite{Wilczek:1987mv}, i.e., the standard electrodynamics
 modified by an additional axion field,  provides a
theoretical framework for a possible violation of parity and Lorentz invariance
 --- the Carroll-Field-Jackiw (CFJ) model \cite{Carroll:1989vb}, \cite{Jackiw:1998js}, \cite{Kostelecky:2002hh}.
The non-abelian extensions of the axion modified electrodynamics for the
Standard Model \cite{Colladay:1998fq}, gravity \cite{Kostelecky:2003fs}, \cite{Jackiw:2003pm} and for supersymmetric models \cite{Belich:2003fa} were worked out. 

The axion itself can be considered as a fundamental field. 
Recently some signals on the axion field observations in PVLAS experiments was reported \cite{Zavattini:2005tm}. 
The new observations, however, do not show the presence of a rotation and ellipticity signals  
and thus stand a strong upper limit on axion  contributions to an optical rotation generated in vacuum by a magnetic field \cite{Zavattini:2005tm}. 

Alternatively, the axion can be viewed as an effective field constructed, for instance,  from  torsion \cite{DeSabbata:1981zk}, 
 \cite{Rubilar:2003uf}, \cite{Itin:2003hr}, \cite{Lammerzahl:2004ww}. 
Astrophysics consequences of such torsion induced axion models are recently studied intensively, see \cite{Preuss:2005ic}, \cite{Maity:2005ah}. 
Moreover, the linear magnetoelectric effects of  $Cr_2O_3$  find a satisfactory explanation in term of a macroscopic axion field, see \cite{Hehl:2007ut}. 
Some mechanisms that actually lead to axion-type modifications of electrodynamics were 
proposed recently in \cite{Klinkhamer:1999zh} and in \cite{Kostelecky:2002ca}. 

  The axion modified electrodynamics is usually formulated by adding a topological Chern-Simons term to the Maxwell Lagrangian \cite{Wilczek:1987mv}, \cite{Carroll:1989vb}, \cite{Jackiw:1998js}. We apply here an alternative  premetric approach to classical electrodynamics \cite{Birkbook}, \cite{serbia}, \cite{Itin:2004za}. In this construction, the axion field  emerges in a natural way as
 an irreducible part of a general constitutive tensor. In the premetric formalism, one starts with two independent antisymmetric fields: the electromagnetic  strength $F_{ij}$ and the excitation field  ${ H}^{ij}$. 
 Here the Roman indices  range from 0 to 3. 
 The Maxwell equations  are given by 
\begin{equation}\label{Max-ten}
\epsilon^{ijkl}F_{ij,k}=0\,, \qquad{ H}^{ij}{}_{,j}={\mathcal J}^i\,,
\end{equation}
where the commas denote the ordinary partial derivatives, $\epsilon_{ijkl}$ is the Levi-Civita permutation tensor normalized
with $\epsilon^{0123}=-\epsilon_{0123}=1$.
The fields $F_{ij}$ and $H^{ij}$ are not independent one on another. For a wide range of physical effect, they can be  assumed to be related by a linear homogeneous constitutive law
\begin{equation}\label{rel}
{ H}^{ij}=\frac 12 \chi^{ijkl}F_{kl}\,.
\end{equation}
By  definition, the constitutive pseudotensor is antisymmetric in two pairs of indices.  
Hence  it  has, in general,  36 independent components. Its irreducible
decomposition  under the group of linear transformations
involves three independent pieces. One of these three pieces is the axion field, which is a
subject of our interest. 
The axion electrodynamics is reinstated in the generic premetric framework by a specialization of the constitutive tensor. It is assumed to be of the following form  \cite{Itin:2004za} 
\begin{equation}\label{ax-const-tensor}
\chi^{ijkl}=\left(g^{ik}g^{jl}-g^{il}g^{jk}\right)\sqrt{-g}+ \psi\epsilon^{ijkl}\,.
\end{equation}
Here $g^{ij}$ is a metric tensor with the signature $\{+,-,-,-\}$ and with the
determinant $g$. 
The axion $\psi$ is a  pseudoscalar field. It is invariant under  transformations of coordinates with a positive determinant and changes its sign under  transformations with a negative determinant. 
In this paper, we restrict to a flat Minkowski spacetime with Cartesian coordinates, so  the square root factor in (\ref{ax-const-tensor}) can be omitted.

In the premetric formalism, the axion field is considered only phenomenologically --- as an intrinsic characteristic of a "media". The dynamics model of the axion field are usually constructed by involving an additional  kinematic term in the Lagrangian, see for instance   \cite{Andrianov:1998ay}. 
The  mathematical methods similar to used here was shown to be useful also in ray
optics applications to GR \cite{Perlick}, \cite{Perlick:2005hz} and in quantum plasmadynamics
\cite{Melrose1}, \cite{Melrose2}.

{\bf 2. "Momentum representation".}
In the premetric electrodynamics the wave propagation is usually studied in the framework of the geometric approximation \cite{Birkbook}. In this case, the variation of the media characteristics (represented by the constitutive tensor $\chi^{ijkl}$) are neglected relative to the change of the wave parameters. Consequently one come to the conclusion that the axion field does not affects the wave  propagation at all \cite{Birkbook}. It is in a contradiction with the standard CFJ electrodynamics predictions \cite{Carroll:1989vb}. The discrepancy is certainly originated in the restrictions of the geometric approximation \cite{Itin:2004za}. 

In order to go beyond the geometric approximation, we start with an ansatz of the form 
 \begin{equation}\label{anzats}
 { F}_{ij}=f_{ij}\,e^{i\phi}\,.
 \end{equation}
 Here the amplitude $f_{ij}$ and the eikonal $\phi$ are arbitrary functions of a point, $i$ is the imaginary unit.  In order to represent by (\ref{anzats}) a wave-type solution, we require the amplitude to vary much slowly then the eikonal function. In other words, we apply a condition 
 \begin{equation}\label{cond}
 \frac {||f_{ij,k}||}{||f_{ij}||} << ||\phi_{,k}|| \,,
 \end{equation}
 where the maximal (matrix) norms are assumed.   We substitute (\ref{anzats}) into equations
 (\ref{Max-ten}), (\ref{rel}) and apply  the condition  (\ref{cond}) to get 
  \begin{equation}\label{eq1a}
 \epsilon^{ijkl}f_{kl}q_{,j}=0\,,
 \end{equation}
 and 
 \begin{equation}\label{eq2a}
 \frac 12 \left(\chi^{ijkl} f_{kl}q_j-i\chi^{ijkl}{}_{,j} f_{kl}\right)=j^i\,.
 \end{equation}
Here the notations $j^i=(-i){\mathcal J}^i\,e^{-i\phi}$ and $q_j=\phi_{,j}$ are used. 
 A most general solution of the linear system (\ref{eq1a}) involves an  arbitrary 
covector $a_k$,  
  \begin{equation}\label{sol1}
 f_{kl}=a_kq_l-a_lq_k\,.
 \end{equation}
 Substituting  into (\ref{eq2a}), we come to  an algebraic system of four linear equations for four components of the  covector  $a_k$
 \begin{equation}\label{main2}
 M^{ik}a_k=j^i\,.
 \end{equation}
Here  the matrix of the system is denoted by 
  \begin{equation}\label{main-matr}
 M^{ik}=\chi^{ijkl}q_lq_j-i\chi^{ijkl}{}_{,j}q_l\,.
 \end{equation}
This  matrix will play a central role in our analysis \cite{Itin:2007av}. 
 After substitution of (\ref{ax-const-tensor}) we write  the $M$-matrix of the axion modified electrodynamics in the form 
\begin{eqnarray}\label{M-matrix}
M^{ij}=(g^{ij}q^2-q^iq^j)+\Pi^{ij}\,,
\end{eqnarray}
where the polarization tensor
\begin{eqnarray}\label{Pi-matrix}
\Pi^{ij}=i\psi_{,k}q_l\epsilon^{ijkl}\,
\end{eqnarray}
 is involved. Due to its  symmetry properties, the $M$-matrix satisfies 
 \begin{equation}\label{M-relations}
M^{ij}q_i=0\,,\qquad M^{ij}q_j=0\,.
\end{equation}
These two relations have a pure physical sense: The former one represents
the charge conservation law, while the latter relation represents  the gauge
invariance of the field equation.

Quite remarkable that the matrix $M$ is Hermitian. Its metric part is a standard real symmetric tensor of vacuum electrodynamics.
The non-metric part represents a polarization tensor $\Pi^{ij}$ 
which is antisymmetric and pure imaginary. 

{\bf 3. Dispersion relation.}
Let us consider  the wave solutions of (\ref{main2}). Four components of the covector $a_k$ satisfy now a homogeneous linear system
\begin{equation}\label{fe-3-new}
M^{ij}a_j=0\,,
\end{equation}
which has a non-zero solution if and only if its determinant equal to
zero.  For the system (\ref{fe-3-new}), this condition holds identically, which can
be seen even without an explicit calculation of the determinant.
 Indeed, the identities
(\ref{M-relations}) express  linear relation between the rows (and between the
columns) of the matrix $M^{ik}$. So this matrix is singular. 
Moreover, (\ref{M-relations}$_b$) also means  that the linear system
(\ref{fe-3-new}) has a non-zero solution of the form $a_j=Cq_j$ 
with an arbitrary constant $C$. This solution is evidently unphysical  since
it can be obtained by a gauge transformation of a trivial zero solution. To describe an
observable  physically meaningful situation, we must have  an additional linear
independent solution of (\ref{fe-3-new}).
A $4\times 4$ linear system (\ref{fe-3-new}) has two linear independent solutions (one for
gauge and one for physics) if and only if its matrix $M^{ij}$ is of rank 2 or
less. An algebraic expression of this requirement is
\begin{equation}\label{Adj1}
A_{ij}=0\,,
\end{equation}
 where $A_{ij}$ is the adjoint matrix.
This matrix is obtained by removing the $i$-th  row and the $j$-th column
from the original matrix $M^{ij}$ . The determinants of the remaining $3\times 3$
matrices are calculated and assembled in a new matrix $A_{ij}$.
 The entries of
the adjoint matrix are expressed via the entries of the matrix $M^{ij}$ as
\begin{equation}\label{Adj2}
A_{ij}=\frac 1{3!}
\epsilon_{ii_1i_2i_3}\epsilon_{jj_1j_2j_3}M^{i_1j_1}M^{i_2j_2}M^{i_3j_3}\,.
\end{equation}
For a matrix satisfying (\ref{M-relations}), the 
 adjoint matrix $A_{ij}$ is symmetric and proportional to the wave covector \cite{itin}
   \begin{equation}\label{Adj2-new}
A_{ij}=\lambda(q)q_iq_j\,.
   \end{equation}
   Since $A_{ij}$ is symmetric and $M^{ij}$ is Hermitian, the adjoint matrix $A_{ij}$ is real. 
   Thus  also the dispersion function $\lambda(q)$ is real. 
Consequently, to guarantee the existence a physically meaningful solution,    we have to require only one real condition
\begin{equation}\label{gen-dis}
\lambda(q)=0\,
\end{equation}
instead of 16 conditions (\ref{Adj1}).
We calculate now the adjoint matrix for the axion modified electrodynamics
model. Substituting (\ref{M-matrix}) into (\ref{Adj2}) and 
calculating in turn  the powers of the
imaginary unit we derive to the adjoint matrix  in the following form
\begin{equation}\label{lambda1}
A_{ij}=-\left[q^4+\left(\psi_{,m}\psi^{,m}\right)q^2-
\left(\psi_{,m}q^m\right)^2\right]q_iq_j\,.
\end{equation}
Thus we come to the known expression of the dispersion relation for the electromagnetic waves in the axion electrodynamics \cite{Jackiw:1998js}, \cite{Itin:2004za}, \cite{serbia}
\begin{equation}\label{disp}
q^4+\left(\psi_{,m}\psi^{,m}\right)q^2-\left(\psi_{,m}q^m\right)^2=0\,.
\end{equation}
Note that this fourth order polynomial  does not admit a covariant factorization to a product of two second order polynomials. It is in spite of the fact that in special coordinates such factorization exists for timelike, spacelike and null covectors $\psi_{,m}$, see \cite{Itin:2007wz}. The expression (\ref{disp}) is not positive defined so the non-birefringence condition \cite{Itin:2005iv} is violated. It is in  correlation  with the result of \cite{Hariton:2006zj}
that typically in axion electrodynamics the Lorentz group is broken down  to the little group associated with the external 4-vector.

{\bf 4. Photon propagator.}
Let us return now to the full inhomogeneous Maxwell equation with a
non-zero current 
 \begin{equation}\label{prop1}
 M^{ij}a_j=j^i\,.
 \end{equation}
The solution of this equation is useful to represent via the Green function or
photon propagator, $D_{ij}(q)$. This matrix is defined in such a way that the
vector
\begin{equation}\label{prop2}
 a_j=-D_{ij}j^i\,
 \end{equation}
is a formal solution of (\ref{prop1}). Due to the gauge invariance and
charge conservation, the propagator, $D_{ij}(q)$,  is defined only up to terms proportional to the wave covector $q_i$, 
\begin{equation}\label{prop3}
 D_{ij}\to D_{ij}+\alpha_iq_j+\beta_jq_i\,.
 \end{equation}
 Here the components of the covectors $\alpha_i$ and $\beta_i$ are arbitrary functions of the
 wave covector.

Since the matrix $M^{ik}$ is singular, the propagator $D_{ij}(q)$ cannot be taken
to be proportional to the  inverse of $M^{ik}$. 
To overcome this obstacle, we use a construction that involves the second adjoint matrix 
\cite{Melrose2}, \cite{Itin:2007av}. This tensor is defined as 
\begin{equation}\label{prop5x}
 B_{ijkl}=\frac 12 \epsilon_{ikm_1n_1}\epsilon_{jlm_2n_2}M^{m_1m_2}M^{n_1n_2}\,.
 \end{equation}
The photon propagator is expressed via this tensor as 
 \begin{equation}\label{result-1}
 D_{ik}=\frac {g^{mn}B_{imnk}}{\lambda q^2}\,.
 \end{equation}
 Calculate the second adjoint matrix for the axion expression (\ref{M-matrix}). 
It is expressed as a sum of three terms
\begin{equation}\label{B-sep}
B_{imnk}={}^{(1)}B_{imnk}+{}^{(2)}B_{imnk}+{}^{(3)}B_{imnk}\,.
 \end{equation}
After removing the gauge terms (\ref{prop3}), the pure metric piece  remains in the form
 \begin{equation}\label{B-1x}
{}^{(1)}B_{imnk}=-q^2g_{ik}q_mq_n\,.
 \end{equation}
 Up to the gauge depending terms, the metric-axion piece is
  \begin{equation}\label{B-2x}
{}^{(2)}B_{imnk}=-iq^2\psi^{,j}\epsilon_{injk}q_m\,.
 \end{equation}
 The third pure axion piece (without the gauge depending terms) takes the form
  \begin{equation}\label{B-3x}
{}^{(3)}B_{imnk}=-\psi_{,i}\psi_{,k}q_mq_n\,.
 \end{equation}
 Thus we have derived the axion modified photon propagator in the following form
  \begin{equation}\label{photon1}
^{(F)}D_{ik}=\frac {q^2g_{ik}+i\psi^{,j}q^n\epsilon_{injk}+\psi_{,i}\psi_{,k}}
{q^4+\left(\psi_{,m}\psi^{,m}\right)q^2-\left(\psi_{,m}q^m\right)^2}\,.
 \end{equation}
 This is an axion generalization of the standard Feynman photon propagator.  Note that in QED one usually multiply $D_{ij}$ by $-i$.
 
Observe that for a constant axion field, the axion modified photon propagator takes the standard vacuum electrodynamics form. 
Moreover, the propagator expression has poles only on the solutions of the dispersion relation $\lambda=0$. Most remarkable that the axion modified photon propagator is hermitian, so the corresponding wave solutions are not damped. The antisymmetric imaginary part appearing in the numerator of (\ref{photon1}) is usual in axion modified models. 

Let us compare the expression (\ref{photon1}) with the standard QED result
\begin{equation}\label{exact-photon1}
\left(D^{-1}\right){\!}^{ik}=\left(\Delta^{-1}\right){\!}^{ik}+\Pi^{ik}\,,
 \end{equation}
 where the free photon propagator $\Delta_{ik}=g_{ik}/q^2$ and the polarization tensor 
$\Pi^{ij}=-i\psi_{,k}q_l\epsilon^{ijkl}$ are involved. 
Recall that this expression is derived by summation of infinite sequence of Feynman diagrams.  
 Multiply the right hand sides of (\ref{photon1}) and (\ref{exact-photon1}) we have 
\begin{eqnarray}\label{compary}
&&^{(F)}D_{ik}\left(D^{-1}\right)^{im}=\frac 1{\lambda} \left(q^2g_{ik}+i\psi^{,j}q^n\epsilon_{injk}+\psi_{,i}\psi_{,k}\right)\times\nonumber\\
&&\qquad\qquad\qquad\qquad\quad\,\,\,\left(q^2g^{im}-i\psi_{,k}q_l\epsilon^{imkl}\right)=\nonumber\\
&&\frac 1{\lambda}
\left(q^4+\left(\psi_{,k}\psi^{,k}\right)q^2-\left(\psi_{,k}q^k\right)^2\right)\delta^i_m=\delta^i_m\,.
\end{eqnarray}
Here we removed the terms proportional to $q^i$ and $q_m$ which are gauge dependent. 
Consequently our expression (\ref{photon1}) is indeed  inverse to (\ref{exact-photon1}). 
 
The Landau photon propagator is derived by removing the transversal terms, ${}^{(L)}D_{ik}q^i={}^{(L)}D_{ki}q^i=0$. It is given by 
 \begin{eqnarray}\label{photon2}
&&{}^{(L)}D_{ik}=\frac 1{\lambda}\Big[g_{ik}-\frac {q_iq_k}{q^2}
+i\psi^{,j}q^n\epsilon_{injk}+\psi_{,i}\psi_{,k}+\nonumber\\
&& \frac{(\psi_{,m}q^m)^2}{q^4}q_iq_k-
\frac{\psi_{,m}q^m}{q^2}(\psi_{,i}q_k+\psi_{,k}q_i)\Big]\,.
\end{eqnarray}

Consider a special version of axion modified electrodynamics with a timelike covector of axion field derivatives --- the CFJ-model \cite{Carroll:1989vb}. 
In this case, the coordinates can be taken in such a way that the axion field derivatives covector is parametrized as $\psi_i=(\mu, 0,0,0)$. Write the wave vector as $q^i=(w,{\mathbf{k}})$. Substituting into (\ref{photon1}) we get 
\[ D_{ik}=\frac 1\lambda\left( \begin{array}{cccc}
w^2-k^2+\mu^2 & 0         & 0         & 0\\
0             & w^2-k^2   & -i\mu k_3 & i\mu k_2\\
0             & i\mu k_3  & w^2-k^2   &-i\mu k_1\\
0             & -i\mu k_2 & i\mu k_1  & w^2-k^2 \end{array} \right)\,.\] 

{\bf 5. Conclusions.}
In this paper, we treat the  axion modified electrodynamics as
 a special case of premetric electrodynamics formalism.
 In this construction, the axion field  emerges as
 an irreducible part of a general constitutive tensor. 
In addition to the known covariant dispersion relation of axion electrodynamics, we have derived a covariant expression of the axion modified photon propagator. 
The axion modified photon propagator in different gauges was constructed recently in  
\cite{Andrianov:1998ay}, \cite{Adam:2001ma}, \cite{Lehnert:2004be}. These expressions share the main properties of (\ref{photon1}) and (\ref{photon2}). In particular, they are represented by a fraction with a numerator which is a quadratic function of the wave covector and with a denominator proportional to the dispersion relation expression. 
Note, however, that the previous considerations are restricted to an axion field with a constant gradient. In our consideration,  this restriction is removed.  
A remarkable fact that the propagator  (\ref{photon1}) is hermitian. Consequently a proper defined energy-momentum tensor has to be conserved without dissipation. This issue lies, however, beyond the scope of the current note.  

{\it Acknowledgments.}
 I am grateful to  Roman Jackiw and Friedrich Hehl for   fruitful
 discussions. I thank the referee for most useful comments and for very relevant references.

 \end{document}